\documentstyle[11pt,newpasp,twoside,epsf]{article}
\def\edcomment#1{\iffalse\marginpar{\raggedright\sl#1\/}\else\relax\fi}
\reversemarginpar

\begin{document}
\title{The Opacity of Spiral Galaxies from Counts of Distant Background Galaxies}
\author{B. W. Holwerda\altaffilmark{1,2}, R. A. Gonz\'alez \altaffilmark{3}, R. J. Allen\altaffilmark{1} and \\ P. C. van der Kruit \altaffilmark{2}}

\affil{(1) Space Telescope Science Institute, Baltimore, MD 21218, USA}
\affil{(2) Kapteyn Institute, Landleven 12, 9747 AD Groningen, The Netherlands}
\affil{(3) Centro de Radioastronom\'{\i}a y Astrof\'{\i}sica, Universidad Nacional Aut\'{o}noma de M\'{e}xico, 58190 Morelia, Michoac\'{a}n, Mexico}

\begin{abstract}
We have applied the "Synthetic Field Method" on a sample of  ~20 nearby galaxies in order to determine the opacity of their disks. We present preliminary results on the radial dependence of cold dust absorption for 3 examples. The spirals NGC4535 and NGC4725 show significant absorption at a half-light radius. UGC2302, a LSB galaxy, shows much less opacity.
\end{abstract}

\section{Introduction}

The Synthetic Field Method (SFM) was developed by Gonz\'alez et al.\ (1998) as a technique for determining the average opacity through the disk of a nearby spiral by counting more distant background galaxies. This number is compared with those of "synthetic fields", a known WFPc2 deep field, dimmed by a certain opacity, added to the foreground galaxy image.

Gonz\'alez et al.\ (2003) examined  the application of the method and concluded that the optimum results with HST/WFPc2 imaging would be obtained on Fornax and Virgo cluster galaxies, with the accuracy degrading for galaxies much closer or more distant.

Our sample was drawn from the HST archives, in part from the Cepheid Distance Scale HST Key project sample.

\section{Results}

We present preliminary results on NGC4535, NGC4725 and UGC2302. The numbers of real and simulated field galaxies were compared in annuli of deprojected radius.  

\subsection{NGC4535 and NGC4725}

NGC4535 is a face-on SABc with a half-light radius of 5.1 kpc at a distance of 16 Mpc. Several spiral arms with bright HII regions are visible in the HST image. NGC4725 is a SABab ringed galaxy at 12 Mpc Its half-light radius is 5.4 kpc. The HST image shows much less structure. The radial opacity profiles of both however are similar (see Figure 1), showing 2.5 magnitudes of extinction in I at the half-light radius, dropping off to 0.5 magnitude. 

\subsection{UGC2302}
 UGC2302 is a LSB at 15 Mpc is a much more compact object. At comparable radii, it does not show as much absorption; less then a magnitude at 5 kpc from the center.

\section{Conclusions}
The uncertainties in these individual profiles are large due to low numbers of field galaxies and an added uncertainty due to their clustering. By averaging a number of fields we will improve on these uncertainties for an average radial opacity model of spiral galaxies.
For different Hubble types, we intend to compare the radial dependence of opacity  
with the HI density profile to obtain cold dust-to-gas ratios.

\begin{figure}
\plottwo{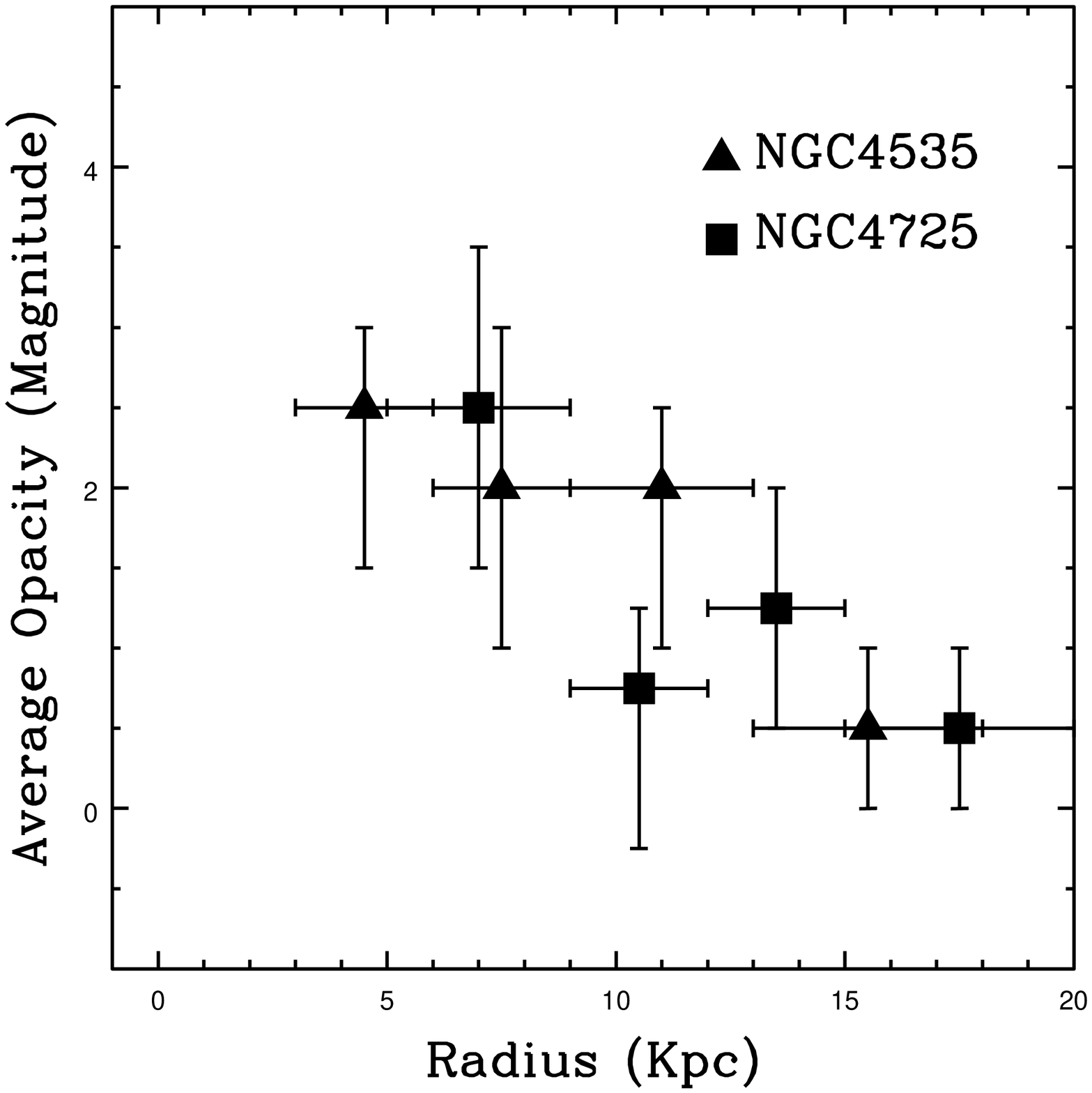}{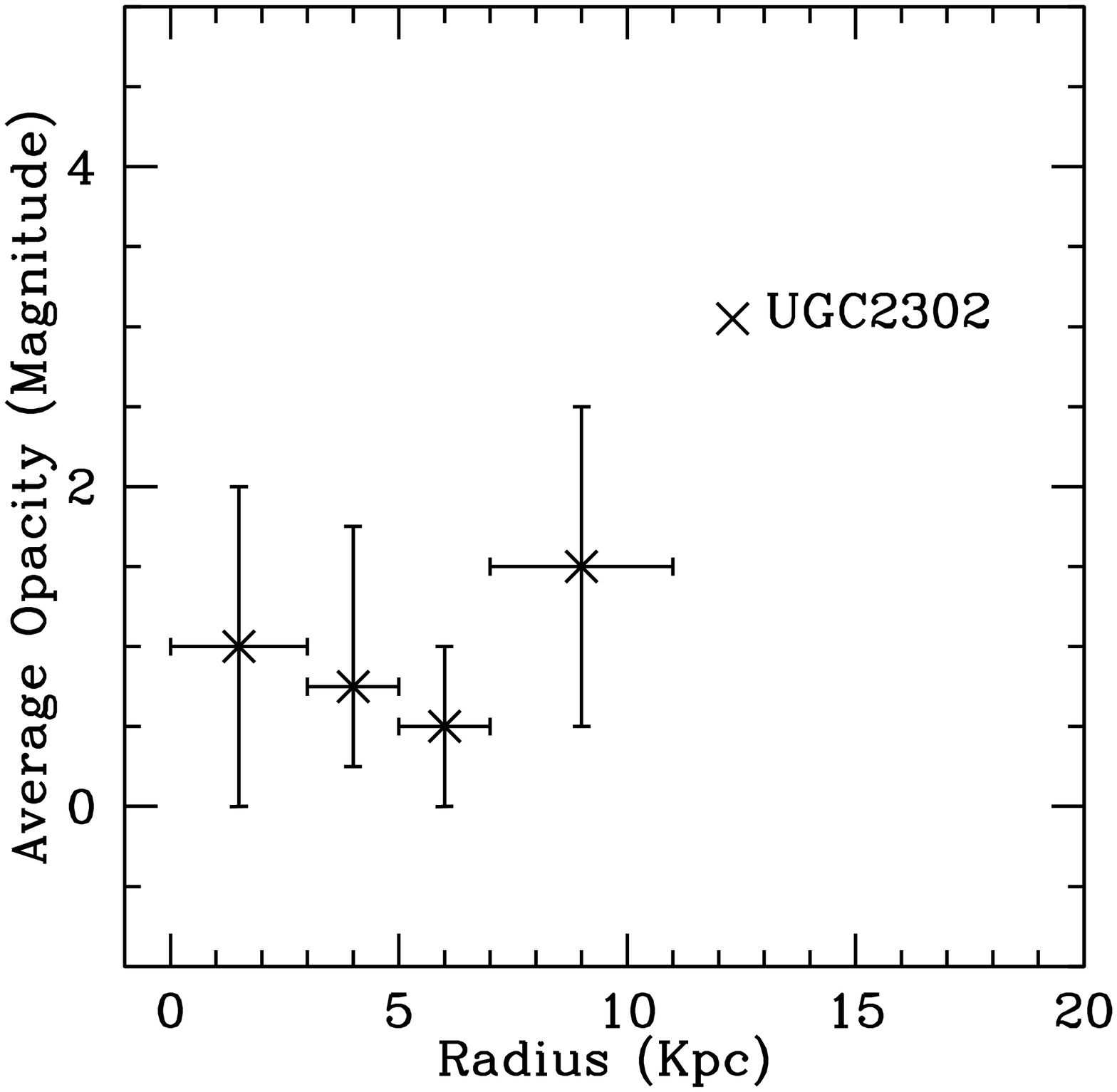}
\caption{The radial opacity measurement from field galaxy counts for NGC4535, NGC4725}
\caption{The radial opacity measurement of UGC2302}
\end{figure}


\begin{references}

\reference Gonz{\' a}lez, R.~A., Allen, R.~J., Dirsch, B., Ferguson, H.~C., Calzetti, D., and Panagia, N. 1998, \apj ~{\bf 506}, 152
\reference Gonz{\' a}lez, R.~A., Loinard, L., Allen, R.~J., and Muller, S.  2003, \aj  ~{\bf 125}, 1182
\reference Holwerda, B.~W., Allen, R.~J., and van der Kruit, P.~C.: 2002, ASP Conf. Ser. 273: 337
\end{references}
\end{document}